\documentclass[prd,onecolumn, nofootinbib, superscriptaddress, 11pt,]{revtex4}  
\usepackage{graphicx}
\usepackage{amsmath,braket}
\usepackage{amsfonts}
\usepackage{amssymb}
\usepackage{bm}
\usepackage{appendix}
\usepackage{mathtools}
\usepackage{comment}
\usepackage{bbold}
\usepackage{color}
\usepackage{slashed}
\usepackage[hyperindex=true, citecolor=green]{hyperref}
\usepackage{subfigure}
\usepackage{setspace}
\usepackage{enumitem}
\usepackage{longtable}
\usepackage{wasysym}
\usepackage[usenames,dvipsnames]{xcolor}
\usepackage{bm}
\usepackage{multirow}
\usepackage{changepage}
\usepackage{kantlipsum}
\usepackage{mathtools}
\usepackage{yfonts}
\usepackage{mathrsfs}
\usepackage[letterpaper, margin=.8in, top=0.85in, bottom=0.85in]{geometry}

\usepackage{tikz}
\usetikzlibrary{decorations.pathmorphing,arrows.meta,bending}

\pdfoutput=1

\newcommand{\e}{\, .}
\newcommand{\ee}{\, ,}

\begin{document}


\hfill{LA-UR-18-29365}

\title{Reparameterization Invariant Operator Basis for NRQED and HQET}

\author{Andrew Kobach}
\affiliation{Theoretical Division, Los Alamos National Laboratory, Los Alamos, NM 87545, USA}
\affiliation{Physics Department, University of California, San Diego, La Jolla, CA 92093, USA}
\author{Sridip Pal}
\affiliation{Physics Department, University of California, San Diego, La Jolla, CA 92093, USA}

\date{\today}

\begin{abstract}
We provide a self-contained discussion of how reparameterization invariance connects a rotationally-invariant heavy particle effective theory with a single heavy fermion to a Lorentz-invariant theory.  Furthermore, using Hilbert-series methods, a Lorentz-invariant operator basis is tabulated, up to and including operators of order $1/M^4$, when the fermion couples to an external $U(1)$ or $SU(3)$ gauge interaction.   
\end{abstract}

\maketitle

\section{\normalsize Introduction}
\label{intro}


Lorentz invariance is intimately connected with existence of antiparticles.  At the same time, one can formulate an effective field theory (EFT) that describes particles which are essentially non-relativistic~\cite{Shifman:1987rj, Isgur:1989vq, Georgi:1990um}.  When considering heavy particles, the energy cost of pair production can be so high that one can integrate out the anti-particles.  Subsequently, both relativistic effects and the effects of anti-particle appear as corrections, i.e., higher-order terms in the EFT, and the relationships between the numerical coefficients of these higher-order terms are due to Lorentz invariance.  The guiding principle to implement the constraints from Lorentz invariance on a heavy-particle effective theory is known as reparameterization invariance (RPI) \cite{Luke:1992cs, Manohar:1997qy, Manohar:2000dt, Brambilla:2003nt, Heinonen:2012km}.  
Imposing invariance under reparameterization can be technically difficult, and this has lead to different perspectives and methods on the topic, e.g., see Refs.~\cite{Luke:1992cs, Manohar:1997qy, Heinonen:2012km}. 

To be concrete, consider the fundamental interactions between gauge field and an elementary fermion $q$ with mass $M$ is governed by the  Lagrangian:
\begin{equation}
\label{pertL}
\mathcal{L} = \overline{q} \left( i\slashed{D} -M \right) q \ee
\end{equation}
where $D$ is the covariant derivative, $D^\mu \equiv \partial^\mu + i g Z A^\mu_a T_a$,  $g$ is the gauge coupling, $gZ$ is the tree level charge of the fermion, the $A^\mu_a$'s are the gauge fields, and the $T_a$'s are the generators of the gauge group.  For other fermionic degrees of freedom $Q$, such as protons, neutrons, or the $b$ quark within a $B$ meson, etc., in general the Lagrangian contains all higher-order, non-renormalizable, operators that are invariants of the Poincar\'{e} group and the gauge group:
\begin{equation}
\label{nonpertL}
\mathcal{L} = \overline{Q} \left( i\slashed{D} -M \right) Q + \frac{a_F g}{4\Lambda} \overline{Q} \sigma_{\alpha\beta}G^{\alpha\beta}Q +\frac{a_D g}{8\Lambda^2} \overline{Q} \gamma_\alpha [D_\beta G^{\alpha\beta}]Q  + \frac{a_C}{\Lambda^2} (\overline{Q}Q)^2  + \cdots \ee
\end{equation}
where the $a$'s are non-perturbative coefficients, $G^{\alpha\beta} \equiv (-i/gZ)[D^\alpha,D^\beta]$, $\sigma^{\alpha\beta} \equiv i[\gamma^\alpha,\gamma^\beta]/2$, and the factors of 4 and 8 in the $a_F$ and $a_D$ operators, respectively, are conventional. Here, $\Lambda$ is the nominal energy scale associated with the effective theory.  We distinguish the gauge coupling $g$ and the tree-level charge $gZ$, since the fermion can have positive or negative charge, e.g, $Z=\pm 1$, or be neutral, where $Z=0$.  The square bracket indicates that the derivative inside only act within the brackets. Here we only include operators that are invariant under parity and time reversal, since the underlying Lagrangian, i.e., Eq.~\eqref{pertL}, is also invariant under these discrete transformations. Due to Lorentz symmetry, the bilinear sector contains no effective operators constructed solely out of covariant derivatives with Lorentz indices symmetric under interchange, otherwise the fermion would not have the relativistic dispersion relation.  Furthermore, in the limit that $g\rightarrow 0$, the fermion does not couple to the external gauge fields, and it is impossible to construct any Lorentz-invariant effective operator which is bilinear in the fermion~\cite{Mannel:2018mqv}.
However, Lorentz symmetry does permit operators quartic in the fermionic fields, even in the $g\rightarrow 0$ limit.  
The operators of different orders of $1/\Lambda$ can, in principle, mix under the renormalization flow.

We focus on a subset of the full Hilbert space of the theory described by the Lagrangian in Eq.~\eqref{nonpertL} that contains only the operators that are bilinear in the fermion, i.e., systems that only contain one fermion:
\begin{equation}
\label{LorentzTheory}
\mathcal{L} = \overline{Q} \left( i\slashed{D} -M  + \frac{a_F g}{4\Lambda} \sigma_{\alpha\beta}G^{\alpha\beta} + \frac{a_Dg}{8\Lambda^2} \gamma_\alpha[D_\beta G^{\alpha\beta}]  + \cdots \right) Q \e
\end{equation}
When the fermion is heavy, one can integrate out the anti-particle component of the relativistic spinor, which generates an infinite number of effective operators, in addition to those already included in Eq.~\eqref{LorentzTheory}.  Doing so will give rise to non-trivial relationships between the Wilson coefficients in the heavy particle effective theory, due to the underlying Lorentz symmetry of the original theory with particles and anti-particles.  We coin this the ``top down,'' perspective.  Later, we will discuss the cases when this fermion is charged under $U(1)$ electromagnetism or $SU(3)$ color, effective theories called NRQED and HQET, respectively.  

There is a second, ``bottom up,'' perspective for these heavy particle effective field theories.  Here, one constructs a theory invariant under only translations and rotations.  The operators that span such a theory are the same as the operators in the heavy particle effective theory after the anti-particles have been integrated out, since integrating out the anti-particles breaks the Lorentz group down to its rotational subgroup: $SO(3,1) \rightarrow SO(3)$.  
In Ref.~\cite{Kobach:2017xkw}, we enumerated an operator basis, invariant under translations, rotations, and the underlying gauge symmetry, for operators bilinear in the fermion, using Hilbert series methods.\footnote{Interestingly, we found that this operator basis can be organized according to irreducible representations of non-relativistic conformal group \cite{Kobach:2018nmt}.}  Such an operator basis can provide the operators for an heavy particle effective field theory, but it does not supply the non-trivial relationships between the Wilson coefficients due to Lorentz symmetry.  Such relationships can be recovered by requiring invariance under an additional transformation, called reparameterization~\cite{Luke:1992cs}.  

We reemphasize that reparameterization invariance can be thought of as a way to implement Lorentz invariance in effective theories with a single heavy degree of freedom, where the anti-particles have already been integrated out~\cite{Luke:1992cs, Manohar:1997qy,Manohar:2000dt, Brambilla:2003nt, Heinonen:2012km}. 
It is relatively straight forward to break Lorentz invariance then integrate out the anti-particles.  But doing it the other way around, i.e., starting with a rotationally-invariant theory and imposing constraints from Lorentz symmetry on a theory with no antiparticles, can be  technically complicated to achieve. 
On a conceptual level, this is what we should expect as a Lorentz boost does mix the particle with the anti-particle. Thus, RPI should somehow know about the existence of anti-particle.  
In Refs.~\cite{Luke:1992cs, Manohar:1997qy, Manohar:2000dt} the form of the RPI transformation changes at higher order in HQET formulation, this in some form implements the information about anti-particle, i.e., the full Lorentz invariance. On other hand, we show that it is possible to formulate a RPI in a way so that the basic transformation by definition knows about the existence of anti-particle. Our work generalizes some of the methods discussed in Refs.~\cite{Luke:1992cs, Manohar:1997qy, Manohar:2000dt}, and provides a complementary viewpoint of how RPI is connected to Lorentz invariance compared to what has been elucidated in Ref.~\cite{Heinonen:2012km}. 

One of the purposes of this paper is to be self-contained, so we revisit both the ``top down'' (Section~\ref{topdown}) and ``bottom up'' (Section~\ref{RPI}) approaches to heavy particle effective field theory with an aim towards to making a rigorous connection between RPI and Lorentz symmetry without spoiling the particle-antiparticle symmetry.  In Section~\ref{RPIBasis}, we explore an immediate corollary of our generalized form of RPI, where operators spanning a reparameterization-invariant theory can be  mapped to those that are manifestly Lorentz-invariant. Thus, we are able to provide with a Lorentz invariant operator basis up to and including order $1/M^{4}$, free of redundancies from integration by parts or equations of motion, for theories with a single fermion, charged under an external $U(1)$ or $SU(3)$ gauge field (Section~\ref{RPIBasis}). Moreover, we explicitly map some of them onto lower dimensional operators in HQET as a proof of our concept along with recovering one of the constraints produced by conventional RPI (see discussion around Eq.~\eqref{prod} and Eq.~\eqref{reprod}).

\section{\normalsize{The ``Top Down'' Approach}}
\label{topdown}
Here, we recapitulate many of the arguments presented in Ref.~\cite{Manohar:2000dt}, using similar notation.
We consider the Lorentz invariant field theory as described by Eq.~\eqref{LorentzTheory}. If the mass $M$ of the particle is heavy compared to all other scales in the system, then its antiparticle can be integrated out, and this induces a set of non-renormalizable effective operators.   To do so, one can factor out the rapidly-oscillating phase of the field, $Q'(x) \equiv e^{iMv\cdot x}Q(x)$, where $v^\mu \equiv (\gamma, \gamma \bm{v})$ is the velocity 4-vector, and $\gamma \equiv 1/\sqrt{1-\bm{v}^2}$, such that $v^2=1$, and the time-ordered two-point correlation function for $Q'$ is:
\begin{eqnarray}
\bra{0} T\overline{Q}'(x)Q'(0) \ket{0} &=& \int \frac{d^4{ p}}{(2\pi)^4} \left( \frac{i}{\slashed{p} - (1-\slashed{v})M + i\epsilon} \right) e^{-i{ p}\cdot { x}} \ee \\
&\simeq& \int \frac{d^4{ p}}{(2\pi)^4} \left(\frac{i}{v\cdot p + i\epsilon} \right)\left(\frac{1+\slashed{v}}{2}\right)e^{-i{ p}\cdot { x}}  + \mathcal{O}\left(\frac{1}{M}\right) \e
\end{eqnarray}
Here, $(1+\slashed{v})/2$ is a projection operator, since $v^2=1$.  In the rest frame, it projects onto the particle component of the field $Q$.  Likewise, $(1-\slashed{v})/2$ is also a projection operator, and in the rest frame, it projects onto the anti-particle component of the field $Q$.  So, the Dirac spinor $Q$ can be decomposed into two components using these projection operators:
\begin{eqnarray}
\label{Qvdef}
Q(x) = e^{-iMv\cdot x} \left[  \underbrace{e^{iMv\cdot x}\left( \frac{1+\slashed{v}}{2}\right)Q(x)}_{\equiv ~Q_v} + \underbrace{e^{iMv\cdot x}\left( \frac{1-\slashed{v}}{2}\right)Q(x)}_{\equiv ~\mathfrak{Q}_v} \right] \ee
\end{eqnarray}
for a general velocity.\footnote{The right-hand side of Eq.~\eqref{Qvdef} does not depend on $v^\mu$.  Therefore, a sum over all $v^\mu$ is the most general expression: $Q(x) =  \sum_v e^{-iMv\cdot x} \left( Q_v(x) + \mathfrak{Q}_v(x) \right)$. 
However, when inserting this definition back into Eq.~\eqref{LorentzTheory}, the Lagrangian will have an overall phase of $e^{\pm iM(v-v')\cdot x}$.   In the $M\rightarrow \infty$ limit, only the sector of the Hilbert space that is not rapidly oscillating in $x$ is the one where $v=v'$, leaving only the sector where all heavy fields have the same velocity~\cite{Georgi:1990um}. We thank E.~Mereghetti for pointing this out. 
}
The Lagrangian in Eq.~\eqref{LorentzTheory} can be rewritten now in terms of $Q_v$ and $\mathfrak{Q}_v$, where  we will let $\Lambda \rightarrow M$, as is conventional in heavy particle effective theory:
\begin{equation}
\mathcal{L} = (\overline{Q}_v + \overline{\mathfrak{Q}}_v)e^{iMv\cdot x} \left( i\slashed{D} -M  + \frac{a_F g}{4M} \sigma_{\alpha\beta}G^{\alpha\beta} + \frac{a_Dg}{8M^2} \gamma_\alpha[D_\beta G^{\alpha\beta}]  + \cdots \right)e^{-iMv\cdot x} ({Q}_v + {\mathfrak{Q}}_v) \e
\end{equation}
For convenience, one can replace operators of the form $\overline{Q}_v i\slashed{D} \mathfrak{Q}_v$ with $\overline{Q}_v i\slashed{D}_\perp \mathfrak{Q}_v$, where $D^\mu_\perp \equiv D^\mu - v^\mu(v\cdot D)$, since $\overline{Q}_v \slashed{v}\mathfrak{Q}_v = 0$, because from Eq.~\eqref{Qvdef}, $\slashed{v}Q_v= Q_v$ and $\slashed{v}\mathfrak{Q}_v = - \mathfrak{Q}_v$:
\begin{equation}
\label{LQvQvfrak}
\mathcal{L} = \overline{Q}_v  \bigg( i\slashed{D} + \frac{a_F g}{4M}  \sigma_{\alpha\beta}G^{\alpha\beta} + \frac{a_Dg}{8M^2} \gamma_\alpha[D_\beta G^{\alpha\beta}] \bigg) Q_v  + \overline{\mathfrak{Q}}_v \left( i\slashed{D}  - 2M \right) \mathfrak{Q}_v  + \overline{Q}_v \left( i\slashed{D}_\perp  + \frac{a_F g}{4M}  \sigma_{\alpha\beta}G^{\alpha\beta}  \right)  \mathfrak{Q}_v + h.c.  +  \cdots \e
\end{equation}
This is the same Lagrangian as Eq.~\eqref{LorentzTheory}.  Some terms have not been included in Eq.~\eqref{LQvQvfrak}, because they contribute at order $1/M^3$ or higher, and our present discussion will be to order $1/M^2$, for the sake of brevity. 

If all operators are bilinear in the heavy fields, the heavy antiparticle $\mathfrak{Q}_v$ can be integrated out by performing the Gaussian integral over $\mathfrak{Q}_v$ in the action.  This is equivalent to solving for the equation of motion for $\mathfrak{Q}_v$:
\begin{equation}
\label{QfrakEOM}
\left( i\slashed{D} - 2M  + \cdots \right)\mathfrak{Q}_v = \left(i \slashed{D}_\perp + \frac{a_F g}{4M}\sigma_{\alpha\beta}G^{\alpha\beta} +  \cdots \right) Q_v  \ee
\end{equation}
and inserting this back into Eq.~\eqref{LQvQvfrak}, noting that $\overline{Q}_v \gamma^\alpha Q_v = v^\alpha\overline{Q}_v Q_v$, and expanding to order $1/M^2$ (after a considerable amount of algebra):
\begin{equation}
\label{RPIHQET}
\mathcal{L} = \overline{Q}_v \bigg[  iv\cdot D - \frac{D^2}{2M} + \frac{(a_F - Z)g}{4M} \sigma_{\alpha\beta}G^{\alpha\beta} - \frac{i(2a_F -Z)g}{8M^2}v_\mu \sigma_{\alpha\beta} \{D_\perp^\alpha, G^{\mu\beta}\} + \frac{a_Dg}{8M^2} v_\alpha [D_{\perp\beta} G^{\alpha\beta}] \bigg] Q_v + \mathcal{O}\left( \frac{1}{M^3} \right) \e
\end{equation}
The non-trivial relationships between the Wilson coefficients of operators at different orders in $1/M$ are due to the underlying theory being Lorentz invariant.  Note that the second operator in Eq.~\eqref{RPIHQET} is not of the form $\overline{Q}_v D_\perp^2 Q$, since we have summed an infinite series of operators to achieve the form $\overline{Q}_v D^2 Q$.  We discuss this point further in Section~\ref{RPI}.
It is interesting to note that the operator $\propto Zg \sigma_{\alpha\beta}G^{\alpha\beta}$ does not depend on a Wilson coefficient and is due to Thomas precession, i.e., it is purely kinematic effect due to the Lorentz group.  This particular form matches onto the Bargmann-Telegdi-Michel equation for the semi-classical motion of a spin-1/2 particle in an external electromagnetic field in the lab frame, which would not have been apparent if one ignored the effective operators in Eq.~\eqref{LorentzTheory}.  

Eq.~\eqref{RPIHQET} is the desired form of the heavy particle effective Lagrangian, subject to external gauge fields. 
The procedure to achieve this form is coined the ``top-down'' approach, and to provide a starting point for this method, we enumerate all Lorentz-invariant operators that span an operator basis in Table~\ref{Table:NRQED} for an external $U(1)$ gauge field, and in Table~\ref{Table:HQET} for an external $SU(3)$ gauge field, up to and including $1/M^4$ operators.  

\section{\normalsize Reparameterization Invariance}
\label{RPI}

A second method by which to derive the non-trivial relationships between Wilson coefficients in heavy particle effective theory is one that begins with a theory invariant under rotations and translations, embeds the rotationally-invariant objects within irreducible representations of the Lorentz group (such that it reduces to the rotationally-invariant theory in the rest frame), and requires  invariance under reparameterization.  We call this the ``bottom-up'' approach, since it does not explicitly use the concept of a Lorentz boost.  These steps yield a Lagrangian of the same form as Eq.~\eqref{RPIHQET}, which we will demonstrate up to and including order $1/M^2$.  
There are other methods by which one can derive the non-trivial relationships between Wilson coefficients (for example, see Refs.~\cite{Sundrum:1997ut, Manohar:1997qy, Brambilla:2003nt, Heinonen:2012km}), which utilize explicit representations of the Lorentz algebra and explicit form of the commutators.  
The method we discuss this this section, based in reparameterization, is a generalization of the one outlined in Refs.~\cite{Luke:1992cs, Manohar:2000dt}.

Reparameterization invariance in heavy particle effective field theories is a consequence of Lorentz invariance, since derivatives in the relativistic theory are split into two operators in the heavy theory.  To illustrate this, consider the theory of a free, relativistic, fermion:
\begin{equation}
\label{freedirac}
\mathcal{L} = \overline{Q} (i \slashed{\partial} - M)Q \e
\end{equation}
Inserting Eq.~\eqref{Qvdef} to rewrite it in terms of $Q_v$ and $\mathfrak{Q}_v$, and integrating out the antiparticle, one obtains:
\begin{eqnarray}
\label{Eq:4.4}
\mathcal{L} = \overline{Q}_v \Bigg[   i v \cdot \partial + i\slashed{\partial}_\perp \frac{1}{\left( iv\cdot \partial + 2M \right) } i\slashed{\partial}_\perp \bigg] Q_v \e
\end{eqnarray}
Expanding in powers of $1/M$:
\begin{eqnarray}
\label{eq14}
\mathcal{L} = \overline{Q}_v \Bigg[   i v \cdot \partial - \frac{\partial_\perp^2}{2M} +  \frac{\partial_\perp^2(iv \cdot  \partial)}{4M^2} - \frac{\partial^2_\perp(iv\cdot \partial)^2}{8M^3} + \cdots \bigg] Q_v \e
\end{eqnarray}
If one inserts the equation of motion for $Q_v$ back into the effective operators in the above Lagrangian, it eliminates all $v$ dependence, and the power series in $1/M$ truncates, resulting in the simple expression:
\begin{eqnarray}
\label{freeL}
\mathcal{L} = \overline{Q}_v \bigg[   i v \cdot \partial - \frac{\partial^2}{2M} \bigg] Q_v \e
\end{eqnarray}
This is the same Lagrangian as Eq.~\eqref{freedirac}, after the antiparticles have been integrated out.\footnote{The passage from Eq.~\eqref{eq14} to Eq.~\eqref{freeL} requires the following to be true:
\begin{eqnarray}
\bigg(- \frac{(iv\cdot \partial)^2}{2M} +  \frac{\partial_\perp^2(iv \cdot  \partial)}{4M^2} - \frac{\partial^2_\perp(iv\cdot \partial)^2}{8M^3} + \cdots \Bigg) Q_v = 0 \e
\end{eqnarray}
This equation can be rewritten as:
\begin{eqnarray}
(iv\cdot \partial )^2 Q_v = (iv\cdot \partial)\left(  \frac{\partial_\perp^2}{2M} - \frac{\partial^2_\perp(iv\cdot \partial)}{4M^2} + \cdots \right)Q_v \ee
\end{eqnarray}
and the two sides of the equation are in fact equal, due to the equation of motion for $Q_v$.  
}
The relative coefficient between the two operators in Eq.~\eqref{freeL} is fixed by the underlying relativistic theory.  The energy-momentum dispersion relation provided by Eq.~\eqref{freeL} is $E = \sqrt{M^2 + (\gamma M {\bm{v}} +{\bf k})^2 } - \gamma M$, where $\bm{v}$ is the 3-velocity, $\gamma \equiv 1/\sqrt{1-\bm{v}^2}$ is the Lorentz factor, and ${\bf k}$ is often called the residual momentum.  This is the relativistic dispersion relation, provided that one identifies the full momentum as $p^\mu = M v^\mu + k^\mu$, and that the energy of the heavy particle has the relativistic mass subtracted.  These relationships are often used as the starting point for heavy particle effective field theory.

Reparameterization invariance is defined as a transformation of the degrees of freedom in Eq.~\eqref{freeL} such that it remains invariant.  This is tantamount to requiring that the relativistic dispersion relation remains intact.  Since the Lagrangians in Eqs.~\eqref{freedirac} and \eqref{freeL} are the same Lagrangian, and since the free Dirac theory in Eq.~\eqref{freedirac} does not depend on the velocity, therefore neither does the heavy effective theory in Eq.~\eqref{freeL}, so a shift in the definition of $v^\mu$ in the effective theory must amount to nothing.    
A sufficient choice for the definition of reparameterization would be to shift the velocity vector $v^\mu \mapsto v^\mu + \varepsilon^\mu/M$~\cite{Luke:1992cs, Manohar:2000dt}.  
Furthermore, because $(1+\slashed{v})/2$ must remain a projection operator, i.e., $[(1+\slashed{v})/2]^n = (1+\slashed{v})/2$ for all $n\in \mathbb{Z}$ where $n>0$, it is necessary to impose constraints on $\varepsilon$ to ensure this. This can be obtained by requiring that $v\cdot \varepsilon = 0$ and terms of order $\mathcal{O}(\varepsilon^2)$ are negligible.   Using the definition in Eq.~\eqref{Qvdef}, one can determine the change in the heavy field, under the shift $v^\mu \mapsto v^\mu + \varepsilon^\mu/M$:
\begin{equation}
\label{RPQv}
Q_{v+\varepsilon/M} = e^{i\varepsilon \cdot x} \left( 1+ \frac{\slashed{\varepsilon}}{2M} \right)Q_v  + e^{i\varepsilon \cdot x} \frac{\slashed{\varepsilon}}{2M}\mathfrak{Q}_v  \e
\end{equation}
This expression is exact and one of the main result of our paper.\footnote{We compare Eq.~\eqref{RPQv} to the one used in Ref.~\cite{Manohar:2000dt}:
\begin{equation}
\label{RPQv2}
Q_{v+\varepsilon/M} = e^{i\varepsilon \cdot x} \left( 1+ \frac{\slashed{\varepsilon}}{2M} \right)Q_v  \ee
\end{equation}
which differs from Eq.~\eqref{RPQv} beginning at order $\mathcal{O}(1/M^2)$.  
}
 Similarly, we have
\begin{equation}
\mathfrak{Q}_{v+\varepsilon/M} = e^{i\varepsilon \cdot x} \left( 1- \frac{\slashed{\varepsilon}}{2M} \right)\mathfrak{Q}_v  -e^{i\varepsilon \cdot x} \frac{\slashed{\varepsilon}}{2M}Q_v \e
\end{equation} 
and we also note that
\begin{equation}\label{RPpsiv}
\left(Q_{v+\varepsilon/M}+\mathfrak{Q}_{v+\varepsilon/M}\right)=e^{i\varepsilon \cdot x} \left(Q_v+\mathfrak{Q}_v\right) \e
\end{equation}
 In the free theory being considered, Eq.~\eqref{RPQv} takes the following form after integrating out the antiparticle:
\begin{equation}
\label{shiftedQ}
Q_{v+\varepsilon/M} = e^{i\varepsilon \cdot x} \left( 1+ \frac{\slashed{\varepsilon}}{2M} + \frac{\slashed{\varepsilon}}{2M} \frac{1}{\left( iv\cdot \partial + 2M \right)} i \slashed{\partial}_\perp \right)Q_v  \e
\end{equation}
After expanding in $1/M$, and using the equation of motion for the heavy field, one can show:
\begin{equation}
\overline{Q}_{v+\varepsilon/M} \bigg[   i \left(v+\frac{\varepsilon}{M} \right) \cdot \partial - \frac{\partial^2}{2M} \bigg] Q_{v+\varepsilon/M}  = \overline{Q}_v \bigg[   i v \cdot \partial - \frac{\partial^2}{2M} \bigg] Q_v \e
\end{equation}
The Lagrangian in Eq.~\eqref{freeL} is invariant under reparameterization, as expected.

Reparameterization invariance supplies a necessary requirement to pass from a rotationally-invariant theory to one that is Lorentz invariant.  Using again the free theory to illustrate this, the most general operator basis for a theory with a free fermion, assuming only rotational and translational invariance is
\begin{equation}
\label{rotfreeL}
\mathcal{L} = \psi^\dagger \bigg\{ i \partial_t + c_2 \frac{\bm{\partial}^2}{2M} + c_4 \frac{\bm{\partial}^4}{8M^3} + \cdots \bigg\} \psi \ee
\end{equation}
where the $c$'s are arbitrary coefficients.  The fermion and derivatives can be embedded within irreducible representations of the Lorentz group:
\begin{equation}
\label{rotfreeLrel}
\mathcal{L} = \overline{Q}_v \bigg\{ i v\cdot \partial - c_2 \frac{\partial_\perp^2}{2M} + c_4 \frac{\partial_\perp^4}{8M^3} + \cdots \bigg\} Q_v \ee
\end{equation}
where $\partial_\perp^\mu \equiv \partial^\mu - v^\mu(v\cdot\partial)$, such that when $v^\mu = (1,0,0,0)$, this reduces to the form of the Lagrangian in Eq.~\eqref{rotfreeL}.  Requiring that the Lagrangian in Eq.~\eqref{rotfreeLrel} is invariant under reparameterization yields $c_2=1$, $c_4=1$, etc.:
\begin{equation}
\label{rotfreeLrel2}
\mathcal{L} = \overline{Q}_v \bigg\{ i v\cdot \partial -  \frac{\partial_\perp^2}{2M} +  \frac{\partial_\perp^4}{8M^3} + \cdots \bigg\} Q_v \e
\end{equation}
When inserting the equation of motion of $Q_v$ back into Eq.~\eqref{rotfreeLrel2} to eliminate all the velocity dependence among the effective operators, the result is the same Lagrangian as Eq.~\eqref{freeL}.

The arguments supporting the existence of reparameterization invariance for a free theory must also carry over to the interacting theory.  A rotationally- and translationally-invariant theory of a two-component Pauli spinor, charged under a gauge group, even under both parity and time reversal, is
\begin{equation}
\label{heqtv0}
\mathcal{L} = \psi^\dagger \bigg\{i D_t + c_2\frac{{\bf D}^2}{2M} + c_F g \frac{\bm{\sigma} \cdot {\bf B}}{2M} +  c_D g \frac{[{\bf D}\cdot {\bf E}]}{8M^2} + i c_S g \frac{\bm{\sigma} \cdot ({\bf D}\times {\bf E} - {\bf E}\times {\bf D})}{8M^2}  \bigg\} \psi + \mathcal{O}\left(\frac{1}{M^3} \right) \e
\end{equation}
Here, $\psi$ is a two-component Pauli spinor, and we have used the convention for the Wilson coefficients in Ref.~\cite{Manohar:1997qy}.  Derivatives acting within square brackets only within those brackets.   This theory can be expressed in an arbitrary frame as:
\begin{equation}
\label{heqtv}
\mathcal{L} = \overline{Q}_v \bigg\{ iv\cdot D - c_2 \frac{D_\perp^2}{2M} - c_F g \frac{\sigma_{\alpha\beta}G^{\alpha\beta}}{4M} - c_D g \frac{[v_\alpha D_{\perp\beta} G^{\alpha\beta} ]}{8M^2} + i c_S g \frac{v_\mu \sigma_{\alpha\beta} \{D_\perp^\alpha, G^{\mu \beta} \}}{8M^2} + \cdots  \bigg\} Q_v + \mathcal{O}\left(\frac{1}{M^3} \right) \e
\end{equation}
Eq.~\eqref{heqtv} reduces to Eq.~\eqref{heqtv0} when $v^\mu = (1,0,0,0)$.    Under reparameterization,  one shifts the velocity by an infinitesimal amount $v^\mu \mapsto v^\mu + \varepsilon^\mu/M$, where $v\cdot \varepsilon = 0$, and the subsequent shift in the heavy field is defined in Eq.~\eqref{RPQv}.   The rotationally-invariant operator basis up to and including order $1/M^4$ operators, invariant under parity, is presented in Refs.~\cite{Kobach:2017xkw, Gunawardana:2017zix}.

Determining the form of $Q_{v+\varepsilon/M}$ after integrating out the antiparticle is non-trivial to an arbitrary order in $1/M$.  If one wishes to impose reparameterization invariance among operators up to and including order $1/M^2$, then Eq.~\eqref{RPQv} takes the form:
\begin{equation}
\label{QceMD}
Q_{v+\varepsilon/M} = e^{i\varepsilon \cdot x} \left( 1+ \frac{\slashed{\varepsilon}}{2M} + \frac{\slashed{\varepsilon}}{2M} \frac{1}{\left( iv\cdot D + 2M \right)} i \slashed{D}_\perp \right)Q_v \e
\end{equation} 
This is identical to the definition of reparameterization of the heavy field in the free theory, under the replacement $\partial \rightarrow D$.  If one wishes to continue imposing reparameterization invariance to high orders in $1/M$, one can begin with the relativistic theory, find the equation of motion for the antiparticle, and insert that relationship into Eq.~\eqref{RPQv}.  

To continue with our discussion, we choose to work to order $1/M^2$, so Eq.~\eqref{QceMD} will serve as the definition of the transformation of the heavy field under reparameterization.  Requiring reparameterization invariance of the Lagrangian yields the following form:
\begin{equation}
\mathcal{L} = \overline{Q}_v \bigg[  iv\cdot D - \frac{D^2}{2M} - \frac{c_Fg}{4M} \sigma_{\alpha\beta}G^{\alpha\beta} + \frac{i(2c_F -Z)g}{8M^2}v_\mu \sigma_{\alpha\beta} \{D_\perp^\alpha, G^{\mu\beta}\} - \frac{c_Dg}{8M^2} v_\alpha [D_{\perp\beta} G^{\alpha\beta}] \bigg] Q_v + \mathcal{O}\left( \frac{1}{M^3} \right) \ee
\end{equation}
which is the same as Eq.~\eqref{RPIHQET}, after identifying that $a_F  = -c_F +Z$ and $a_D = -c_D$.  These are the same results found in Refs.~\cite{Manohar:1997qy, Heinonen:2012km}.  The relationships from reparameterization invariance (or, rather, Lorentz invariance) between the Wilson coefficients up to and including $1/M^3$ for HQET and NRQED can be founds in Refs.~\cite{Manohar:1997qy} and \cite{Heinonen:2012km}, respectively, and some of the relationships for NRQED at $1/M^4$ can be found in \cite{Hill:2012rh}.  These results utilized different methods than the ones discussed here.

\section{\normalsize Operator Basis for Reparameterization-invariant NRQED and HQET}
\label{RPIBasis}

A method to ensure a reparameterization-invariant operator basis is to construct the effective Lagrangian out of bilinear operators
\begin{equation}
\label{RPIinvL}
\mathcal{L}_\text{eff} = \sum_k \overline{\Psi}_v \mathcal{O}_k \Psi_v \ee
\end{equation}
where $\Psi_v$ and $\mathcal{O}_k$ themselves transform covariantly under reparameterization.  This method is discussed in Refs.~\cite{Luke:1992cs, Manohar:1997qy, Heinonen:2012km,Hill:2012rh}.  We present here a general version of this method, using the definitions provided in Section~\ref{intro}.  Specifically, the heavy field $\Psi_v$ is defined as
\begin{equation}
\Psi_v \equiv  Q_v + \mathfrak{Q}_v \ee
\end{equation} 
which, using Eq.~\eqref{RPpsiv}, transforms under reparameterization as:
\begin{equation}
\Psi_{v+\varepsilon/M} = e^{i\varepsilon\cdot x} \Psi_v \e
\end{equation}
The operator $\mathcal{O}_k$ is constructed out of Dirac matrices, field strength tensors (both of which are invariant under reparameterization), and covariant derivatives, $\mathfrak{D}_\mu$ defined as:
\begin{equation}
i\mathfrak{D}_\mu \equiv iD_\mu  + Mv_\mu \ee
\end{equation}
where $D_\mu$ is the gauge covariant derivative, such that $i\mathfrak{D}_\mu \Psi_v $ transforms covariantly under reparameterization:
\begin{equation}
i\mathfrak{D}_\mu \Psi_v ~ \mapsto ~ (iD_\mu  + Mv_\mu + \varepsilon_\mu) \Psi_{v+\varepsilon/M} = e^{i\varepsilon\cdot x} i\mathfrak{D}_\mu \Psi_v \e
\end{equation}
The operator $\mathcal{O}_k$ cannot be constructed out of $v^\mu$, since it does not transform linearly under reparameterization.   Therefore, any bilinear operator in Eq.~\eqref{RPIinvL} will be invariant under reparameterization.

One may proceed in this manner, defining an operator basis for the theory defined in Eq.~\eqref{RPIinvL}, free from redundancies associated with integration by parts and equations of motion when calculating $S$-matrix elements~\cite{Politzer:1980me, Georgi:1991ch}.  Before doing so, it is interesting to remember the definition in Eq.~\eqref{Qvdef}:
\begin{equation}
Q = e^{-iMv\cdot x} \Psi_v \ee
\end{equation}
where, again, $Q$ is the Dirac spinor.  So, bilinear operators built out of objects that transform covariantly under reparameterization can be written solely in terms of objects in the Lorentz-invariant theory.  For example:
\begin{equation}
\overline{\Psi}_v i\mathfrak{D}_\mu \Psi_v = \overline{Q} iD_\mu Q \e
\end{equation}
Any reparameterization-invariant operator in Eq.~\eqref{RPIinvL} can be rewritten as a Lorentz-invariant one by making the trivial replacement $\Psi_v \rightarrow Q$ and $\mathfrak{D}_\mu \rightarrow D_\mu$.

We continue by defining the operator basis for an explicitly Lorentz-invariant theory, since the nomenclature is more conventional.  To aid in the construction of an operator basis, we use Hilbert-series methods, as laid out in Refs.~\cite{Feng:2007ur, Jenkins:2009dy, Hanany:2010vu, Lehman:2015via, Lehman:2015coa, Henning:2015daa, Henning:2015alf, Henning:2017fpj}.  To begin by defining the objects out of which we will construct singlets of the Lorentz group and the gauge group.  When in three spatial dimensions, is most natural to use the local isomorphism $SO(3,1) \simeq SU(2)_L \times SU(2)_R$, due to the simplicity of the $SU(2)$ algebra.   See Table~\ref{irreps} for the irreducible representations of the Lorentz and gauge group, i.e., $U(1)$ and $SU(3)$, for the objects out of which the effective Lagrangian is built.

\begin{center}
\begin{table}[h!]
\begin{tabular}{ | c || c | c | c |}
\hline
Symbol  & $SU(2)_L$ & $SU(2)_R$ & $U(1)$ \\ \hline \hline
$\psi$ & {\bf 2} & {\bf 1} & 1 \\ \hline
$\psi^\dagger$ & {\bf 1} & {\bf 2} &  -1 \\ \hline  
$\psi^c$ & {\bf 2} & {\bf 1} &  -1 \\ \hline 
$\psi^{c\dagger}$ & {\bf 1} & {\bf 2} &  1 \\ \hline
$F_L$ & {\bf 3} & {\bf 1} &  0 \\ \hline 
$F_R$ & {\bf 1} & {\bf 3} &  0 \\ \hline 
$\mathcal{D}$ & {\bf 2} & {\bf 2} &  0 \\ \hline 
\end{tabular} 
\hspace{0.5in}
\begin{tabular}{ | c || c | c | c |}
\hline
Symbol  & $SU(2)_L$ & $SU(2)_R$ & $SU(3)$ \\ \hline \hline
$\psi$ & {\bf 2} & {\bf 1} & {\bf 3} \\ \hline
$\psi^\dagger$ & {\bf 1} & {\bf 2} &  ${\bf \bar{ 3}}$ \\ \hline  
$\psi^c$ & {\bf 2} & {\bf 1} &  ${\bf { \bar{3}}}$ \\ \hline 
$\psi^{c\dagger}$ & {\bf 1} & {\bf 2} &  ${\bf { 3}}$ \\ \hline
$G_L$ & {\bf 3} & {\bf 1} &  ${\bf 8}$ \\ \hline 
$G_R$ & {\bf 1} & {\bf 3} &  ${\bf 8}$ \\ \hline 
$\mathcal{D}$ & {\bf 2} & {\bf 2} &  ${\bf 1}$ \\ \hline 
\end{tabular} 

\caption{Left:~The irreducible representations of the Lorentz and gauge group for the objects out of which our effective Lagrangian is built.  The normalization  of the $U(1)$ charge is moot, since we are only making singlets in the bilinear sector. Right:~Same as the left-hand table, but for an $SU(3)$ gauge group.}
\label{irreps}
\end{table}
\end{center}
%

Exploring first the case of an external $U(1)$ gauge, we define Hilbert series as 
\begin{equation}
\label{HSQED}
HS = \oint [d\alpha]_{SU(2)_L} \oint [d\beta]_{SU(2)_R} \oint [dz]_{U(1)} ~PE_\psi ~PE_{\psi^\dagger} ~PE_{\psi^c} ~PE_{\psi^{c\dagger}} ~PE_{F_L}~ PE_{F_L}  \ee
\end{equation}
where 
\begin{eqnarray}
PE_{\psi^\star} &\equiv& \exp\left[\sum_{n=1}^\infty \frac{(-1)^{(n+1)}(\psi^\star)^n}{n} P(\mathcal{D}^n,\alpha^n,\beta^n) \chi_{\psi^\star}(\mathcal{D}^n, \alpha^n,\beta^n,z^n) \right] \ee \\
PE_{F_\star} &\equiv& \exp\left[\sum_{n=1}^\infty \frac{(F_\star)^n}{n} P(\mathcal{D}^n,\alpha^n,\beta^n) \chi_{F_\star}(\mathcal{D}^n, \alpha^n,\beta^n,z^n) \right] \ee \\
P(\mathcal{D},\alpha,\beta) &\equiv& \frac{1}{(1-\mathcal{D}\alpha\beta)(1-\mathcal{D}\alpha/\beta)(1-\mathcal{D}\beta/\alpha)(1-\mathcal{D}/\alpha\beta)} \ee \\
\oint [d\alpha]_{SU(2)_L} & \equiv& \oint_{|\alpha|=1} \frac{d\alpha}{2\alpha} \left(1- \alpha^2 \right) \left(1 - \frac{1}{\alpha^2} \right) \ee \\
\oint [d\beta]_{SU(2)_R} & \equiv& \oint_{|\beta|=1} \frac{d\beta}{2\beta} \left(1- \beta^2 \right) \left(1 - \frac{1}{\beta^2} \right) \ee \\ 
\oint [dz]_{U(1)} & \equiv& \oint_{|z|=1} \frac{dz}{z} \ee  
\end{eqnarray}
where $\psi^\star$ stands for $\psi, \psi^\dagger, \psi^c$ or $\psi^{c\dagger}$, and $F_\star$ stands for $F_L$ or $F_R$.  The characters $\chi$ for the Weyl fermions contain a subtraction due to the choice of basis that operators of the form $\slashed{D} Q$ are ignored, since they can be related to other operators in the basis via the equations of motion for $Q$:
\begin{eqnarray}
\chi_\psi(\mathcal{D} , \alpha ,\beta ,z ) &\equiv& z\left( \alpha + \frac{1}{\alpha} - \mathcal{D} \left( \beta + \frac{1}{\beta} \right) \right) \ee \\
\chi_{\psi^\dagger}(\mathcal{D} , \alpha ,\beta ,z ) &\equiv& \frac{1}{z} \left( \beta + \frac{1}{\beta} - \mathcal{D} \left( \alpha + \frac{1}{\alpha} \right) \right) \ee \\
\chi_{\psi^c}(\mathcal{D} , \alpha ,\beta ,z ) &\equiv& \frac{1}{z} \left( \alpha + \frac{1}{\alpha} - \mathcal{D} \left( \beta + \frac{1}{\beta} \right) \right) \ee \\
\chi_{\psi^{c\dagger}}(\mathcal{D} , \alpha ,\beta ,z ) &\equiv& z \left( \beta + \frac{1}{\beta} - \mathcal{D} \left( \alpha + \frac{1}{\alpha} \right) \right) \e
\end{eqnarray}
In a full quantum field theory, $D_\alpha F^{\alpha\beta}$ can be related to other operators in the Hilbert space via the equations of motion for $F^{\alpha\beta}$.  However, since we are working only in the single-particle sector, it is possible that the fermions in our sector can respond to external gauge fields, so in our case $D_\alpha F^{\alpha\beta}$ cannot be ignored, in general.  All the while, we must maintain the Bianchi identity, $D_\alpha \widetilde{F}^{\alpha\beta}=0$.  Because $F_L = F + i \widetilde{F}$ and $F_R = F - i \widetilde{F}$, we choose to subtract operators with $D {F}_R$, but not those with $D {F}_L$, so therefore the characters for the gauge field with these relations are:
\begin{eqnarray}
\chi_{F_L}(\alpha,\beta,z) &\equiv& \alpha^2 + 1 + \frac{1}{\alpha^2} \ee \\
\chi_{F_R}(\alpha,\beta,z) &\equiv& \beta^2 + 1 + \frac{1}{\beta^2}  - \mathcal{D}\left(\alpha + \frac{1}{\alpha} \right) \left(\beta + \frac{1}{\beta} \right) + \mathcal{D}^2  \e
\end{eqnarray}
After Taylor expanding the integrand in Eq.~\eqref{HSQED} to second order in the fermions, and performing the integrals over the unit circles, the Hilbert series at each mass dimension is:
\begin{eqnarray}
HS_{d=5} &=& \psi \psi^c F_L + \psi^\dagger \psi^{c\dagger}F_R \ee \\ 
HS_{d=6} &=& \psi^c \psi^{c\dagger} F_L \mathcal{D} + \psi \psi^{\dagger} F_L \mathcal{D}  \ee \\
HS_{d=7} &=& \psi \psi^c F_L \mathcal{D}^2 + \psi^\dagger \psi^{c\dagger} F_L \mathcal{D}^2 + \psi \psi^c F_L^2 + \psi \psi^c F_R^2 + \psi^\dagger \psi^{c\dagger} F_L^2 + \psi^\dagger \psi^{c\dagger} F_R^2 \ee \\
HS_{d=8} &=& \psi^c \psi^{c\dagger} F_L \mathcal{D}^3 + \psi \psi^{\dagger} F_L \mathcal{D}^3 + \psi^c \psi^{c\dagger}F_L^2 \mathcal{D} +  \psi \psi^{\dagger}F_L^2 \mathcal{D} + 2\psi \psi^\dagger F_L F_R \mathcal{D} + 2\psi^c \psi^{c\dagger} F_L F_R \mathcal{D}  \e 
\end{eqnarray}
At this level, the Hilbert series does not say how to contract indices, and it includes all operators of any charge under the parity ($P$) and  time reversal ($T$). Using the Hilbert series output as a guide, we explicitly construct the operators, contracting Lorentz indices by hand, and categorize them by their charge under $P$ and $T$, as done in Table~\ref{Table:NRQED}. The operator basis for an external electromagnetic interaction, for example, would be spanned by only operators even  under both $P$ and $T$.

Here, we pause for a brief aside to illustrate how we go from the Hilbert series output to operators that are listed Tables~\ref{Table:NRQED} and~\ref{Table:HQET}.  For example, we can consider Hilbert series output for $d=5$:
\begin{align}
HS_{d=5} = \psi \psi^c F_L + \psi^\dagger \psi^{c\dagger}F_R \e  
\end{align}
Since the fundamental objects are two-component spinors, one can construct two Hermitian operators, invariant under $CPT$, by contracting the spinor indices: 
\begin{align}
\mathcal{O}_1&\equiv (\psi^c)^{\alpha} (F_L)_{\alpha}{}^{\beta}\ \psi_{\beta}+(\psi^{\dagger})_{\dot{\alpha}} (F_R)^{\dot{\alpha}}{}_{\dot{\beta}}\ (\psi^c{}^{\dagger})^{\dot{\beta}}\,,\\
\mathcal{O}_2&\equiv i\left[(\psi^c)^{\alpha} (F_L)_{\alpha}{}^{\beta}\ \psi_{\beta}-(\psi^{\dagger})_{\dot{\alpha}}(F_R)^{\dot{\alpha}}{}_{\dot{\beta}}\ (\psi^c{}^{\dagger})^{\dot{\beta}} \right]\,.
\end{align}
These operators can be recast using the familiar vector indices:
\begin{align}
\mathcal{O}_1&=\overline{\Psi}\sigma_{\mu\nu}F^{\mu\nu}\Psi\,,\\
\mathcal{O}_2&=\overline{\Psi}\sigma_{\mu\nu}\widetilde{F}^{\mu\nu}\Psi\,.
\end{align}
where $\Psi$ is related to $\psi_{\alpha}$ and $(\psi^c{}^{\dagger})^{\dot{\alpha}}$ in Weyl basis:
\begin{equation}
\Psi=\begin{pmatrix}
\psi_{\alpha}\\
(\psi^{c}{}^{\dagger})^{\dot{\alpha}}
\end{pmatrix}\,,\quad \overline{\Psi}=\begin{pmatrix}
(\psi^c)^{\alpha} \,, & (\psi^{\dagger})_{\dot{\alpha}} 
\end{pmatrix} \e
\end{equation}
Among the two operators $\mathcal{O}_1$ and $\mathcal{O}_2$, we see that only $\mathcal{O}_{1}$ is $P$ and $T$ even. It is interesting to note the comparison to the Lagrangian with only heavy particles.  To do so, one can switch to the Dirac basis, which separates the particle and anti-particle:
\begin{align}
\label{prod}
\mathcal{O}_{1}=\underbrace{\psi^\dagger\ [{\sigma}\cdot {\bf B}]\ \psi}_{\text{heavy-particle operator}} +\ \text{terms involving anti-particle}\,, \quad\ \text{where}\quad \psi=\left(\frac{1+\gamma^{0}}{2}\right)\Psi\,.
\end{align}
which is precisely the operator in Eq.~\eqref{heqtv0}, modulo a multiplicative constant.  

In practice, we do not go through this exercise for all operators appears as output of the Hilbert series.  Instead, we use the Hilbert series as a guide for how many singlet operators there are with the indicated degrees of freedom.  It turns out that, up to and including dimension 8, the operator basis can be expressed as Hermitian operators with derivatives only acting on the field strength tensors, and not the fermions, as demonstrated in Table~\ref{Table:NRQED}.  One might wonder at this point how our choice of basis with no derivatives acting on the fermionic degrees of freedom compares to the HQET operators as written down in Ref.~\cite{Manohar:1997qy}, which contains operators that do.  The key observation is that these terms come about via integrating out the antiparticle component of terms appearing at lower dimension in Table~\ref{Table:NRQED}.  For simplicity, let us elucidate on how this happens in NRQED.   For example, to explain the $c_M$ terms, we look at the operator $ \mathcal{O}_3\equiv\overline{\Psi}\gamma_\mu  [\partial_\nu F^{\mu\nu}]\Psi$ appearing at dimension $6$ in the Table~\ref{Table:NRQED}. Expanding this out in the Dirac basis, we see that $\mathcal{O}_3$ contains a piece that mixes the particle ($\psi_1$) and the anti-particle ($\psi_2$) fields, given by
\begin{align}
\mathcal{O}_3 \ni \psi_1^\dagger (\sigma\cdot {\bf j})\psi_2 + h.c.\,,
\end{align}
where $({\bf j})^{i}\equiv j^{i}= \partial_{\nu}F^{\nu i}$. This mixed piece can be recast in terms of heavy fields $Q_{v=0}\equiv Q$ and $\mathfrak{Q}_{v=0}\equiv \mathfrak{Q}$ in the rest frame as $Q^{\dagger}(\sigma\cdot {\bf j})\mathfrak{Q}+h.c.,$ by using using Eq.~\eqref{Qvdef}.  If one integrates out the antiparticle piece $\mathfrak{Q}$ using $\mathfrak{Q}= \frac{i}{2m}\sigma\cdot{\bf D} Q$, it produces a contribution to the $c_M$ term, as evident from the following expression: 
\begin{align}
\mathcal{O}_3=Q^{\dagger}(\sigma\cdot {\bf j})\mathfrak{Q}+h.c \ni -\frac{i}{2m} Q^{\dagger} \left(\bf D\cdot[\bf D \times \bf B]+[\bf D \times\bf B]\cdot \bf D\right)Q \,.
\end{align}
We also note that particle piece in $\mathcal{O}_3$ reproduces the $c_D$ term:
\begin{align}
\mathcal{O}_3\ni \psi_1^\dagger (\bf D\cdot \bf E)\psi_1 \,.
\end{align}
The above observation is consistent with the relation 
\begin{equation}
\label{reprod}
2c_M=c_F - c_D \e
\end{equation} In fact, one can verify that the $c_F$ contribution also comes about by integrating out the antiparticle. This explicitly shows how operators at different orders in $1/M$ mix upon imposing reparameterization invariance and integrating out the anti-particle.

We now repeat the exercise for an external $SU(3)$ gauge field.  Similar as before, the Hilbert series is defined to be:
\begin{equation}
\label{HSHQET}
HS = \oint [d\alpha]_{SU(2)_L} \oint [d\beta]_{SU(2)_R} \oint [dz_1,dz_2]_{SU(3)} ~PE_\psi ~PE_{\psi^\dagger} ~PE_{\psi^c} ~PE_{\psi^{c\dagger}} ~PE_{G_L}~ PE_{G_L}  \ee
\end{equation}
where 
\begin{equation}
\oint [dz_1,dz_2]_{SU(3)} \equiv \oint_{|z_1|,|z_2|=1}\frac{dz_1 dz_2}{6z_1 z_2} (1-z_1z_2) \left(1 - \frac{z_1^2}{z_2} \right)\left(1 - \frac{z_2^2}{z_1} \right) \left(1 - \frac{1}{z_1 z_2} \right)\left(1 - \frac{z_1}{z_2^2} \right)\left(1 - \frac{z_2}{z_1^2} \right) \ee
\end{equation}
the definition for the $PE$'s are the same as in the $U(1)$ case, but now the characters involve color charge:
\begin{eqnarray}
\chi_\psi(\mathcal{D} , \alpha ,\beta ,z_1,z_2 ) &\equiv& \chi^{SU(3)}_{\bf 3}(z_1, z_2) \left( \alpha + \frac{1}{\alpha} - \mathcal{D} \left( \beta + \frac{1}{\beta} \right) \right) \ee \\
\chi_{\psi^\dagger}(\mathcal{D} , \alpha ,\beta ,z_1,z_2 ) &\equiv& \chi^{SU(3)}_{\bf \bar{3}}(z_1, z_2) \left( \beta + \frac{1}{\beta} - \mathcal{D} \left( \alpha + \frac{1}{\alpha} \right) \right) \ee \\
\chi_{\psi^c}(\mathcal{D} , \alpha ,\beta ,z_1,z_2 ) &\equiv& \chi^{SU(3)}_{\bf \bar{3}} \left( \alpha + \frac{1}{\alpha} - \mathcal{D} \left( \beta + \frac{1}{\beta} \right) \right) \ee \\
\chi_{\psi^{c\dagger}}(\mathcal{D} , \alpha ,\beta ,z_1,z_2 ) &\equiv& \chi^{SU(3)}_{\bf{3}} \left( \beta + \frac{1}{\beta} - \mathcal{D} \left( \alpha + \frac{1}{\alpha} \right) \right) \ee \\
\chi_{G_L}(\alpha,\beta,z_1,z_2) &\equiv& \chi^{SU(3)}_{\bf{8}} \left( \alpha^2 + 1 + \frac{1}{\alpha^2}\right) \ee \\
\chi_{G_R}(\alpha,\beta,z_1,z_2) &\equiv&\chi^{SU(3)}_{\bf{8}} \left( \beta^2 + 1 + \frac{1}{\beta^2}  - \mathcal{D}\left(\alpha + \frac{1}{\alpha} \right) \left(\beta + \frac{1}{\beta} \right) + \mathcal{D}^2 \right)  \e
\end{eqnarray}
where
\begin{eqnarray}
\chi^{SU(3)}_{\bf 3}(z_1, z_2)  &\equiv& z_1 + \frac{z_2}{z_1} + \frac{1}{z_2} \ee \\
\chi^{SU(3)}_{\bf \bar{3}}(z_1, z_2)  &\equiv& z_2 + \frac{z_1}{z_2} + \frac{1}{z_1}\ee \\
\chi^{SU(3)}_{\bf 8}(z_1, z_2)  &\equiv& z_1 z_2 + \frac{z_2^2}{z_1} + \frac{z_1^2}{z_2} + 2 + \frac{z_1}{z_2^2} + \frac{z_2}{z_1^2} + \frac{1}{z_1z_2} \e 
\end{eqnarray}
After Taylor expanding the integrand in Eq.~\eqref{HSHQET} to second order in the fermions, and performing the integrals, the Hilbert series is
\begin{eqnarray}
HS_{d=5} &=& \psi \psi^c G_L + \psi^\dagger \psi^{c\dagger}G_R \ee \\ 
HS_{d=6} &=& \psi^c \psi^{c\dagger} G_L \mathcal{D} + \psi \psi^{\dagger} G_L \mathcal{D}  \ee \\
HS_{d=7} &=& \psi \psi^c G_L \mathcal{D}^2 + \psi^\dagger \psi^{c\dagger} G_L \mathcal{D}^2 + 3\psi \psi^c G_L^2 +2 \psi \psi^c G_R^2 + 2\psi^\dagger \psi^{c\dagger} G_L^2 + 3\psi^\dagger \psi^{c\dagger} G_R^2 \ee \\
HS_{d=8} &=& \psi^c \psi^{c\dagger} G_L \mathcal{D}^3 + \psi \psi^{\dagger} G_L \mathcal{D}^3 + 4\psi^c \psi^{c\dagger}G_L^2 \mathcal{D} +  4\psi \psi^{\dagger}G_L^2 \mathcal{D} + 6\psi \psi^\dagger G_L G_R \mathcal{D} + 6\psi^c \psi^{c\dagger} G_L G_R \mathcal{D}  \nonumber \\
&& \qquad +~ \psi^c \psi^{c\dagger} G_R^2 \mathcal{D} + \psi \psi^{\dagger} G_R^2 \mathcal{D} \e
\end{eqnarray}
With this output as an aid, we contract Lorentz indices by hand, and categorize all operators by their charge under $P$ and $T$, as shown in Table~\ref{Table:HQET}.  The operator basis when the gauge theory is $SU(3)$ color is even under $P$ and $T$.

\begin{table}[h!tbp]
\centering
{
\begin{tabular}{| c || c | c | c | c |}
\hline
Order & $P$ even, $T$ even & $P$ even, $T$ odd & $P$ odd, $T$ even & $P$ odd, $T$ odd \\ \hline \hline 

$M^{-1}$& $  \sigma_{\mu\nu} F^{\mu\nu} $ &  & & $ \sigma_{\mu\nu} \widetilde{F}^{\mu\nu} $  \\  \hline

$M^{-2}$ & $  \gamma_\mu  [\partial_\nu F^{\mu\nu}]$ &  & $  \gamma_\mu \gamma^5  [\partial_\nu F^{\mu\nu}]$ & \\  \hline

\multirow{3}{*}{$M^{-3}$} & $   F_{\mu\nu} F^{\mu\nu} $ & & & $ i\gamma^5  F_{\mu\nu} F^{\mu\nu} $ \\
					     & $  i\gamma^5 F_{\mu\nu} \widetilde{F}^{\mu\nu} $ & & & $   F_{\mu\nu} \widetilde{F}^{\mu\nu} $ \\   
					     & $  \sigma_{\mu\nu}  [\partial^2 F^{\mu\nu}]  $ & & & $  \sigma_{\mu\nu}  [\partial^2 \widetilde{F}^{\mu\nu}]  $ \\ \hline

\multirow{4}{*}{$M^{-4}$} & $  \gamma_\mu  [\partial_\nu \partial^2 F^{\mu\nu}]$ & $  \gamma^\alpha  F_{\alpha\nu} [\partial_\mu F^{\mu\nu}]$ & $  \gamma_\mu \gamma^5  [\partial_\nu \partial^2 F^{\mu\nu}]$ & $  \gamma^\alpha \gamma^5  F_{\alpha\nu} [\partial_\mu F^{\mu\nu}]$  \\
					     & $  \gamma^\alpha \gamma^5  \widetilde{F}_{\alpha\nu} [\partial_\mu F^{\mu\nu}]$ & & $  \gamma^\alpha  \widetilde{F}_{\alpha\nu} [\partial_\mu F^{\mu\nu}]$  &     \\   
					     & $\gamma_\mu \gamma^5  {F}_{\alpha\nu} [\partial^\alpha \widetilde{F}^{\mu\nu}]$ & & $  \gamma_\mu  {F}_{\alpha\nu} [\partial^\alpha \widetilde{F}^{\mu\nu}]$ &  \\ \hline					     
					     		  
\end{tabular}
}
\caption{A basis of Hermitian, Lorentz-invariant, effective operators in a relativistic theory of a single fermion, subject to an external $U(1)$ gauge interaction, categorized by their charge under parity $(P)$ and time reversal ($T$) transformations, up to and including dimension 8.   The operators $\mathcal{O}$ listed in this table should be understood as sandwiched between two Dirac spinors, i.e., $\overline{Q}\mathcal{O}Q$.  The square brackets indicate that the derivatives act only on the object within the square brackets. While this is explicitly a Lorentz-invariant theory, it can be rewritten as a reparameterization-invariant theory by making the replacements $Q \rightarrow \Psi_v$ and $D_\mu \rightarrow \mathfrak{D}_\mu$.  See Section~\ref{RPIBasis} for definitions and details. }
\label{Table:NRQED}
\end{table}

\begin{table}[h!tbp]
\centering
{
\begin{tabular}{| c || c | c | c | c |}
\hline
Order & $P$ even, $T$ even & $P$ even, $T$ odd & $P$ odd, $T$ even & $P$ odd, $T$ odd \\ \hline \hline 

$M^{-1}$& $  \sigma_{\mu\nu}  G_a^{\mu\nu}T^a  $ &  & & $ \sigma_{\mu\nu} \widetilde{G}_a^{\mu\nu}T^a $  \\  \hline

$M^{-2}$ & $  \gamma_\mu [D_\nu G^{\mu\nu}]_a T^a $ &  & $  \gamma_\mu \gamma^5   [D_\nu G^{\mu\nu}]_a T^a $ & \\  \hline

\multirow{5}{*}{$M^{-3}$} & $G_{\mu\nu a} G^{\mu\nu}_b \delta^{ab}$  & $i\gamma^5 G_{\mu\nu a} \widetilde{G}^{\mu\nu}_b T_c f^{abc}$ & $G_{\mu\nu a} \widetilde{G}^{\mu\nu}_b T_c f^{abc}$ & $ i\gamma^5 G_{\mu\nu a} G^{\mu\nu}_b \delta^{ab}$  \\
					     & $G_{\mu\nu a} G^{\mu\nu}_b T_c d^{abc}$ & & & $i\gamma^5 G_{\mu\nu a} G^{\mu\nu}_b T_c d^{abc}$ \\   
					     & $i\gamma^5 G_{\mu\nu a} \widetilde{G}^{\mu\nu}_b \delta^{ab}$  & & & $G_{\mu\nu a} \widetilde{G}^{\mu\nu}_b \delta^{ab}$  \\ 
					     & $i\gamma^5 G_{\mu\nu a} \widetilde{G}^{\mu\nu}_b T_c d^{abc}$ & & & $G_{\mu\nu a} \widetilde{G}^{\mu\nu}_b T_c d^{abc}$ \\ 
					     & $\sigma_{\mu\nu}  [D^2 G^{\mu\nu}]_a T^a $ & & & $\sigma_{\mu\nu}  [D^2 \widetilde{G}^{\mu\nu}]_a T^a $ \\
					     \hline

\multirow{8}{*}{$M^{-4}$} & $\gamma_\mu [D_\nu D^2 G^{\mu\nu}]_a T^a$ & $\gamma^\alpha G_{\alpha\nu a} [D_\mu G^{\mu\nu}]_b \delta^{ab}$  & $\gamma^5 \gamma_\mu [D_\nu D^2 G^{\mu\nu}]_a T^a$  & $\gamma^\alpha \gamma^5  G_{\alpha\nu a} [D_\mu G^{\mu\nu}]_b \delta^{ab}$   \\
					     & $\gamma^\alpha G_{\alpha\nu a} [D_\mu G^{\mu\nu}]_b T_c f^{abc}$ & $\gamma^\alpha G_{\alpha\nu a} [D_\mu G^{\mu\nu}]_b T_c d^{abc}$ & $\gamma^\alpha \gamma^5  G_{\alpha\nu a} [D_\mu G^{\mu\nu}]_b T_c f^{abc}$ & $\gamma^\alpha \gamma^5  G_{\alpha\nu a} [D_\mu G^{\mu\nu}]_b T_c d^{abc}$ \\   
					     & $ \gamma^\alpha \gamma^5\widetilde{G}_{\alpha\nu a} [D_\mu G^{\mu\nu}]_b \delta^{ab}$ & $ \gamma^\alpha \gamma^5 \widetilde{G}_{\alpha\nu a} [D_\mu G^{\mu\nu}]_b T_c f^{abc}$ & $\gamma^\alpha \widetilde{G}_{\alpha\nu a} [D_\mu G^{\mu\nu}]_b \delta^{ab}$ & $\gamma^\alpha \widetilde{G}_{\alpha\nu a} [D_\mu G^{\mu\nu}]_b T_c f^{abc}$ \\
					     & $\gamma^\alpha \gamma^5  \widetilde{G}_{\alpha\nu a} [D_\mu G^{\mu\nu}]_b T_c d^{abc}$ &  $\gamma_\mu G_{\alpha\nu a} [D^\alpha \widetilde{G}^{\mu\nu}]_b T_c f^{abc}$ & $\gamma^\alpha \widetilde{G}_{\alpha\nu a} [D_\mu G^{\mu\nu}]_b T_c d^{abc}$ & $\gamma_\mu G_{\alpha\nu a} [D^\alpha \widetilde{G}^{\mu\nu}]_b T_c f^{abc}$ \\
					     & $\gamma_\mu \gamma^5 G_{\alpha\nu a} [D^\alpha \widetilde{G}^{\mu\nu}]_b \delta^{ab}$ & $\gamma^\alpha [D_\alpha G_{\mu\nu a}G^{\mu\nu}]_b T_c d^{abc}$ & $\gamma_\mu G_{\alpha\nu a} [D^\alpha \widetilde{G}^{\mu\nu}]_b \delta^{ab}$ & $\gamma^\alpha \gamma^5 [D_\alpha G_{\mu\nu a}G^{\mu\nu}]_b T_c d^{abc}$  \\
					     & $\gamma_\mu \gamma^5 G_{\alpha\nu a} [D^\alpha \widetilde{G}^{\mu\nu}]_b T_c d^{abc}$ & & $\gamma_\mu G_{\alpha\nu a} [D^\alpha \widetilde{G}^{\mu\nu}]_b T_c d^{abc}$ &  \\ 
					     & $\gamma^\alpha \gamma^5 [D_\alpha G_{\mu\nu a}\widetilde{G}^{\mu\nu}]_b T_c d^{abc}$ & & $\gamma^\alpha [D_\alpha G_{\mu\nu a}\widetilde{G}^{\mu\nu}]_b T_c d^{abc}$ & \\
					     \hline 
					     		  
\end{tabular}
}
\caption{The same as Table~\ref{Table:NRQED}, but for an external $SU(3)$ gauge interaction. Here, the $SU(3)$ color indices are suppressed, and Roman letters $a, b, c,$ etc., are the indices associated with the eight generators $T^a$ of $SU(3)$.  }
\label{Table:HQET}
\end{table}

\section{\normalsize Discussion and Summary}

This work is the culminating step in our program for constructing invariant operator basis in heavy particle effective theories. In Ref.~\cite{Kobach:2017xkw}, we developed and employed a Hilbert-series method to construct and enumerate an operator basis in a rotationally-invariant theory of a single fermion in an external gauge field. In Ref.~\cite{Kobach:2018nmt} we showed that the operator basis in Ref.~\cite{Kobach:2017xkw} is spanned by scalar primaries of the non-relativistic conformal group.   This present article provides a discussion and generalization of a particular point of view that makes the connection between a rotationally-invariant theory and one that is Lorentz invariant.  An important link between these two theories is requiring reparameterization invariance, which relates operators appearing in different orders in $1/M$.  

Reparameterization invariance is a necessary consequence of Lorentz symmetry in effective theories with a single heavy degree of freedom, where the anti-particles have been integrated out~\cite{Manohar:1997qy,Manohar:2000dt, Brambilla:2003nt, Heinonen:2012km}. 
While Lorentz symmetry necessarily requires the existence of anti-particles, it may be surprising on a face value that requiring invariance under reparameterization yields the same constraints as Lorentz symmetry, since the original effective field theory is formulated only with reference to particle degrees of freedom.  It is clear, however, from Eq.~\eqref{RPQv}, that the reparameterization transformation picks up components from the anti-particle degrees of freedom, in such a way that respects Lorentz symmetry.  In this sense, the reparameterized shift in the velocity, i.e., $v^\mu \mapsto v^\mu + \varepsilon^\mu/M$, where $v\cdot \epsilon = 0$, could be interpreted as a infinitesimal, norm-preserving, Lorentz boost in an arbitrary frame.   However, nowhere does one necessarily require invoking the algebraic concept of a Lorentz boost in order to derive the constraints from reparameterization invariance (for other examples, see Refs.~\cite{Sundrum:1997ut, Manohar:1997qy, Brambilla:2003nt, Heinonen:2012km}).  

We revisit the unambiguous ``top-down'' approach, which begins with a Lorentz-invariant theory, and explicitly integrates out the anti-particles, as discussed in Section~\ref{topdown}.  A second ``bottom-up'' approach, as discussed in Section~\ref{RPI}, begins with a translationally- and rotationally-invariant theory, and requires reparameterization invariance.  This second method has been the cause of some debate in the literature. We present a general treatment of this method, extending the work in Ref.~\cite{Manohar:2000dt}, including an exact expression for the reparameterized heavy field in Eq.~\eqref{RPQv}.  We show that both the ``top-down'' and ``bottom-up'' methods produce the same theory up to and including order $1/M^2$.  These methods can be used to determine the heavy-particle Lagrangian to higher orders in $1/M$, though with significant increase in algebraic complexity, the results of which are discussed in Refs.~\cite{Manohar:1997qy, Heinonen:2012km, Hill:2012rh}.

The exact expression for the reparameterized heavy field in Eq.~\eqref{RPQv} involves of both the particle and anti-particle, which upon integrating out the anti-particles, and expanding to fixed order in $1/M$, becomes the one generally used in the literature, e.g., Ref.~\cite{Manohar:2000dt}.  The novelty associated with this is that we are able to establish an one-to-one correspondence between a theory that is explicitly invariant under reparameterization and a theory that is Lorentz invariant, as discussed in Section~\ref{RPIBasis}.

Because of this one-to-one correspondence between operators that are invariant under reparameterization and ones that are Lorentz invariant, we tabulate an operator basis, using the Lorentz-invariant notation.  
We use Hilbert series methods, with a similar setup as in Refs.~\cite{Henning:2015daa, Henning:2015alf,Henning:2017fpj}, but with the modification that one of the gauge fields is in a long representation of the conformal group, since we are restricted to the Hilbert space with only one matter degree of freedom.  While the Hilbert series provides the number of  invariant operators given the field content, we contract indices by hand, and categorize the Hermitian operators by their charges under the discrete transformations of parity and time reversal, as tabulated in Table~\ref{Table:NRQED} for NRQED and Table~\ref{Table:HQET} for HQET.  It is interesting to note that this relativistic theory spanned by bilinear operators in a fermion, subject to external gauge fields is also the starting point for SCET~\cite{Bauer:2000ew, Bauer:2000yr, Beneke:2002ph}.

\acknowledgements
We are grateful for conversations with Brian Henning, Shauna Kravec, Aneesh Manohar, Duff Neill, and Emanuele Mereghetti.  The work of AK and SP is supported in part by DOE grant \#DE-SC0009919, and the work of AK is also supported  in part by the US DOE Office of Nuclear Physics and by the LDRD program at Los Alamos National Laboratory.

\bibliographystyle{JHEP}

\bibliography{bib}{}

\providecommand{\href}[2]{#2}\begingroup\raggedright\begin{thebibliography}{10}

\bibitem{Shifman:1987rj}
M.~A. Shifman and M.~B. Voloshin, {\it {On Production of $D$ and $D^*$ Mesons
  in $B$ Meson Decays}},  {\em Sov. J. Nucl. Phys.} {\bf 47} (1988) 511. [Yad.
  Fiz.47,801(1988)].

\bibitem{Isgur:1989vq}
N.~Isgur and M.~B. Wise, {\it {Weak Decays of Heavy Mesons in the Static Quark
  Approximation}},  {\em Phys. Lett.} {\bf B232} (1989) 113--117.

\bibitem{Georgi:1990um}
H.~Georgi, {\it {An Effective Field Theory for Heavy Quarks at Low-energies}},
  {\em Phys. Lett.} {\bf B240} (1990) 447--450.

\bibitem{Manohar:1997qy}
A.~V. Manohar, {\it {The HQET/NRQCD Lagrangian to order $\alpha / m^3$}},  {\em
  Phys. Rev.} {\bf D56} (1997) 230--237,
  [\href{http://arxiv.org/abs/hep-ph/9701294}{{\tt hep-ph/9701294}}].

\bibitem{Manohar:2000dt}
A.~V. Manohar and M.~B. Wise, {\it {Heavy quark physics}},  {\em Camb. Monogr.
  Part. Phys. Nucl. Phys. Cosmol.} {\bf 10} (2000) 1--191.

\bibitem{Brambilla:2003nt}
N.~Brambilla, D.~Gromes, and A.~Vairo, {\it {Poincare invariance constraints on
  NRQCD and potential NRQCD}},  {\em Phys. Lett.} {\bf B576} (2003) 314--327,
  [\href{http://arxiv.org/abs/hep-ph/0306107}{{\tt hep-ph/0306107}}].

\bibitem{Heinonen:2012km}
J.~Heinonen, R.~J. Hill, and M.~P. Solon, {\it {Lorentz invariance in heavy
  particle effective theories}},  {\em Phys. Rev.} {\bf D86} (2012) 094020,
  [\href{http://arxiv.org/abs/1208.0601}{{\tt arXiv:1208.0601}}].

\bibitem{Luke:1992cs}
M.~E. Luke and A.~V. Manohar, {\it {Reparametrization invariance constraints on
  heavy particle effective field theories}},  {\em Phys. Lett.} {\bf B286}
  (1992) 348--354, [\href{http://arxiv.org/abs/hep-ph/9205228}{{\tt
  hep-ph/9205228}}].

\bibitem{Mannel:2018mqv}
T.~Mannel and K.~K. Vos, {\it {Reparametrization Invariance and Partial
  Re-Summations of the Heavy Quark Expansion}},  {\em JHEP} {\bf 06} (2018)
  115, [\href{http://arxiv.org/abs/1802.09409}{{\tt arXiv:1802.09409}}].

\bibitem{Kobach:2017xkw}
A.~Kobach and S.~Pal, {\it {Hilbert Series and Operator Basis for NRQED and
  NRQCD/HQET}},  {\em Phys. Lett.} {\bf B772} (2017) 225--231,
  [\href{http://arxiv.org/abs/1704.00008}{{\tt arXiv:1704.00008}}].

\bibitem{Kobach:2018nmt}
A.~Kobach and S.~Pal, {\it {Conformal Structure of the Heavy Particle EFT
  Operator Basis}},  {\em Phys. Lett.} {\bf B783} (2018) 311--319,
  [\href{http://arxiv.org/abs/1804.01534}{{\tt arXiv:1804.01534}}].

\bibitem{Sundrum:1997ut}
R.~Sundrum, {\it {Reparameterization invariance to all orders in heavy quark
  effective theory}},  {\em Phys. Rev.} {\bf D57} (1998) 331--336,
  [\href{http://arxiv.org/abs/hep-ph/9704256}{{\tt hep-ph/9704256}}].

\bibitem{Gunawardana:2017zix}
A.~Gunawardana and G.~Paz, {\it {On HQET and NRQCD Operators of Dimension 8 and
  Above}},  {\em JHEP} {\bf 07} (2017) 137,
  [\href{http://arxiv.org/abs/1702.08904}{{\tt arXiv:1702.08904}}].

\bibitem{Hill:2012rh}
R.~J. Hill, G.~Lee, G.~Paz, and M.~P. Solon, {\it {NRQED Lagrangian at order
  $1/M^4$}},  {\em Phys. Rev.} {\bf D87} (2013) 053017,
  [\href{http://arxiv.org/abs/1212.4508}{{\tt arXiv:1212.4508}}].

\bibitem{Politzer:1980me}
H.~D. Politzer, {\it {Power Corrections at Short Distances}},  {\em Nucl.
  Phys.} {\bf B172} (1980) 349--382.

\bibitem{Georgi:1991ch}
H.~Georgi, {\it {On-shell effective field theory}},  {\em Nucl. Phys.} {\bf
  B361} (1991) 339--350.

\bibitem{Feng:2007ur}
B.~Feng, A.~Hanany, and Y.-H. He, {\it {Counting gauge invariants: The
  Plethystic program}},  {\em JHEP} {\bf 03} (2007) 090,
  [\href{http://arxiv.org/abs/hep-th/0701063}{{\tt hep-th/0701063}}].

\bibitem{Jenkins:2009dy}
E.~E. Jenkins and A.~V. Manohar, {\it {Algebraic Structure of Lepton and Quark
  Flavor Invariants and CP Violation}},  {\em JHEP} {\bf 10} (2009) 094,
  [\href{http://arxiv.org/abs/0907.4763}{{\tt arXiv:0907.4763}}].

\bibitem{Hanany:2010vu}
A.~Hanany, E.~E. Jenkins, A.~V. Manohar, and G.~Torri, {\it {Hilbert Series for
  Flavor Invariants of the Standard Model}},  {\em JHEP} {\bf 03} (2011) 096,
  [\href{http://arxiv.org/abs/1010.3161}{{\tt arXiv:1010.3161}}].

\bibitem{Lehman:2015via}
L.~Lehman and A.~Martin, {\it {Hilbert Series for Constructing Lagrangians:
  expanding the phenomenologist's toolbox}},  {\em Phys. Rev.} {\bf D91} (2015)
  105014, [\href{http://arxiv.org/abs/1503.07537}{{\tt arXiv:1503.07537}}].

\bibitem{Lehman:2015coa}
L.~Lehman and A.~Martin, {\it {Low-derivative operators of the Standard Model
  effective field theory via Hilbert series methods}},  {\em JHEP} {\bf 02}
  (2016) 081, [\href{http://arxiv.org/abs/1510.00372}{{\tt arXiv:1510.00372}}].

\bibitem{Henning:2015daa}
B.~Henning, X.~Lu, T.~Melia, and H.~Murayama, {\it {Hilbert series and operator
  bases with derivatives in effective field theories}},  {\em Commun. Math.
  Phys.} {\bf 347} (2016), no.~2 363--388,
  [\href{http://arxiv.org/abs/1507.07240}{{\tt arXiv:1507.07240}}].

\bibitem{Henning:2015alf}
B.~Henning, X.~Lu, T.~Melia, and H.~Murayama, {\it {2, 84, 30, 993, 560, 15456,
  11962, 261485, ...: Higher dimension operators in the SM EFT}},  {\em JHEP}
  {\bf 08} (2017) 016, [\href{http://arxiv.org/abs/1512.03433}{{\tt
  arXiv:1512.03433}}].

\bibitem{Henning:2017fpj}
B.~Henning, X.~Lu, T.~Melia, and H.~Murayama, {\it {Operator bases,
  $S$-matrices, and their partition functions}},  {\em JHEP} {\bf 10} (2017)
  199, [\href{http://arxiv.org/abs/1706.08520}{{\tt arXiv:1706.08520}}].

\bibitem{Bauer:2000ew}
C.~W. Bauer, S.~Fleming, and M.~E. Luke, {\it {Summing Sudakov logarithms in $B
  \rightarrow X(s \gamma)$ in effective field theory}},  {\em Phys. Rev.} {\bf
  D63} (2000) 014006, [\href{http://arxiv.org/abs/hep-ph/0005275}{{\tt
  hep-ph/0005275}}].

\bibitem{Bauer:2000yr}
C.~W. Bauer, S.~Fleming, D.~Pirjol, and I.~W. Stewart, {\it {An Effective field
  theory for collinear and soft gluons: Heavy to light decays}},  {\em Phys.
  Rev.} {\bf D63} (2001) 114020,
  [\href{http://arxiv.org/abs/hep-ph/0011336}{{\tt hep-ph/0011336}}].

\bibitem{Beneke:2002ph}
M.~Beneke, A.~P. Chapovsky, M.~Diehl, and T.~Feldmann, {\it {Soft collinear
  effective theory and heavy to light currents beyond leading power}},  {\em
  Nucl. Phys.} {\bf B643} (2002) 431--476,
  [\href{http://arxiv.org/abs/hep-ph/0206152}{{\tt hep-ph/0206152}}].

\end{thebibliography}\endgroup

\end{document}